\documentclass[%
 preprint,
 amsmath,amssymb,
 aps,
]{revtex4-1}

\usepackage{rotating}

\usepackage{epsfig}
\usepackage{epstopdf}
\usepackage{graphicx}
\usepackage{bm}
\usepackage{float}
  
\begin{document}

\pagenumbering{arabic}

\title{Strong pinning in the hole-doped pnictide superconductor La$_{0.34}$Na$_{0.66}$Fe$_2$As$_2$}
\author{Shyam Sundar$^{1,\ddagger}$, S. Salem-Sugui, Jr.$^1$, A.D. Alvarenga$^2$, M. M. Doria$^1$, Yanhong Gu$^{3,4}$, Shiliang Li$^{3,4,5}$, Huiqian Luo$^{3,5}$ and L. Ghivelder$^1$}
\affiliation{$^1$Instituto de Fisica, Universidade Federal do Rio de Janeiro, 21941-972 Rio de Janeiro, RJ, Brazil}
\affiliation{$^2$Instituto Nacional de Metrologia Qualidade e Tecnologia, 25250-020 Duque de Caxias, RJ, Brazil}
\affiliation{$^3$Beijing National Laboratory for Condensed Matter Physics, Institute of Physics, Chinese Academy of Sciences, Beijing  100190, China}
\affiliation{$^4$University of Chinese Academy of Sciences, Beijing 100049, China}
\affiliation{$^5$Songshan Lake Materials Laboratory, Dongguan, Guangdong 523808, China}
\affiliation{$^\ddagger$Present address: Department of Physics, Simon Fraser University, Burnaby, British Columbia, Canada V5A 1S6}

\email{shyam.phy@gmail.com}



\begin{abstract}

We present magnetization studies as a function of time, temperature and magnetic field for $H$ $\parallel$ c-axis, in a hole-doped pnictide superconductor, La$_{0.34}$Na$_{0.66}$Fe$_2$As$_2$, with, $T_c$ $\approx$ 27 K. The obtained vortex phase-diagram shows that the magnetic irreversibility line is very close to the mean-field superconducting transition line, similar to the low $T_c$ superconductors, evidencing a strong pinning behavior. The irreversibility line does not follow a power law behavior with ($T_c$-$T$), however, it is well described using an expression developed in the literature considering the effect of disorder in the system. The critical current density estimated using the Bean's critical-state model is found to be of the order of 10$^5$ A/cm$^2$ below 12 K in the limit of zero magnetic field. A plot of the normalized pinning force density as a function of the reduced magnetic field at different temperatures shows a good scaling and the analysis suggests that the vortex pinning is due to normal point like pinning centers. The temperature dependence of the critical current density suggests that the pinning due to the variation in charge carrier mean free path alone is not sufficient to explain the experimental data. Magnetic relaxation rate as a function of temperature and magnetic field is also studied.

\end{abstract}

\vspace{2pc}

\maketitle

\section{Introduction}

Superconductivity in Fe-pnictides is a field of great interest for fundamental science as well as for technological advancement \cite{fa11, pen18, hid18, pal15}. Since the advent of iron-based superconductors (IBS), many new superconductors have been discovered in different families of this interesting class of high $T_c$ superconducting compounds \cite{hid18}. Among the different families of IBS, the 122-class is more interesting in-terms of technological purpose and also due to the availability of good quality sizable single crystals \cite{pal15, hid18, asw12, abd14, mah15, gri13}. Recently,  a new platform in the 122-pnictide family was discovered with a chemical formula (La$_{0.5-x}$Na$_{0.5+x}$)Fe$_2$As$_2$ ((La,Na)-122 family) \cite{2, yan15}. It is an interesting and unique member of the 122-family, which allows the investigation of the electron-hole asymmetry, because in this system the doping occurs at the La-Na site with no changing in the Fe-As layers, as opposed to the doping in other 122-pnictide systems \cite{1, 2, yan15}.     

Vortex dynamics and pinning mechanism in IBS is quite interesting due to its salient features \cite{baruch2, FeNi, AsP, pra13} such as, low Ginzburg number ($G_i$) \cite{ele17}, small anisotropy \cite{yua09}, high intergrain connectivity \cite{wei12}, high upper critical field \cite{sen08} and moderate $T_c$ \cite{ren08}. These properties are vital for technological applications of type-II superconductors \cite{pal15, hid18}. In-spite of low Ginzburg number ($G_i$) the vortex phase-diagram of IBS and low-$T_c$ superconductors (LTS) are quite different in the sense that the upper critical-field, $H_{c2}$(T), and the irreversibility line, $H_{irr}$(T), are separated by a broad vortex-liquid region in IBS \cite{shi18}, whereas, in LTS, no appreciable vortex-liquid region exist. Our motivation in the present study is to explore the vortex-dynamics in this recently discovered 122-type of IBS \cite{1}.  

In this work, we report a study of vortex dynamics in a single crystal of the hole-doped La$_{0.34}$Na$_{0.66}$Fe$_2$As$_2$ superconductor using isofield temperature dependence of the magnetization, $M$($T$), isothermal magnetic field dependence of the magnetization, $M$($H$), and magnetic relaxation, $M$($t$) measurements for $H$ $\parallel$ c-axis. The obtained vortex phase-diagram shows that the irreversibility line, $H_{irr}$, is very close to the mean field $T_c$($H$)-line, similar to the behavior in LTS. Critical current density, $J_c$, and the pinning mechanism are analysed using the models developed by Griessen et al \cite{wen} and Dew-Hughes \cite{dew} respectively. Magnetic relaxation rate as a function of temperature and magnetic field is discussed.  

\section{Experimental}

The hole doped La$_{0.34}$Na$_{0.66}$Fe$_2$As$_2$ crystal was obtained accidentally when growing LaFeAsO single crystal with NAs and NaF flux. Details of the crystal growth are given in Ref. \cite{1}. The obtained La$_{0.34}$Na$_{0.66}$Fe$_2$As$_2$ crystals were characterized using different experimental techniques (see Ref.\cite{1}), which confirm the good quality of the crystals. These crystals have the same structure as of A$_e$Fe$_2$As$_2$ (A$_e$ = alkaline earth metal, e.g. Ca, Sr, Ba) with La and Na occupying the A$_e$ site. A sample from the same batch as in Ref. \cite{1} with mass $m$= 0.1348 mg and dimensions 1.54 mm $\times$ 1.52 mm $\times$ 0.016 ($\pm$ 0.002) mm was used in the present study. The magnetic measurements were made using a Quantum Design vibrating sample magnetometer (VSM) built in a 9 T physical property measurement system (PPMS), with magnetic fields applied parallel to the $c$-axis of the sample. $M$($T$) measurement at $H$ = 10 Oe shows a superconducting transition $T_c$ $\approx$ 27 K and a transition width $\Delta{T_c}$ $\approx$ 4 K. All magnetic measurements were obtained after a standard zero field cooled (ZFC) procedure. Isothermal $M$($H$) were measured with a ramp field d$H$/d$t$ = 50 Oe/sec, and obtained in five quadrants. Isofield $M$($T$) were continuously measured during a slow heating (and cooling in the case of field cooled (FC) curves) of d$T$/d$t$ = 0.3 K/sec. $M$($t$) curves were obtained for a span time of 1.5 hours at various temperatures with $H$ = 30 kOe and at various magnetic fields at $T$ = 16 K.

\section{Results and discussion}

Figure 1a shows selected isofield $M(T)$ curves obtained in both ZFC and FC modes. The ZFC and FC curves separate just below the temperature region where the magnetization appears to be flat, with a subtle diamagnetic inclination (normal background). Figure 1b shows a detail of the $M(T)$ curve obtained for $H$ = 1 T, where the background selected in the normal region follows the expression $M_{back}$ = $a$($H$) - $b$($H$)$T$, where $a$($H$) and $b$($H$) $>$ 0 are constants for each magnetic field. For curves with $H$ $>$ 3 kOe, $a$($H$) is negative, which is due to the contribution from sample holder. This background contribution was observed for all $M$($T$) curves and subtracted in the following analysis. The inset of Fig. 1b shows  a detail of the main plot near the transition which clearly shows that it is difficult to resolve the difference between $T_c$($H$), the mean field transition temperature defined as the temperature for which magnetization become diamagnetic with decreasing temperature, and $T_{irr}$, which marks the irreversible temperature at which the ZFC and FC curves separate. This behavior has been observed for all $M$($T$) curves. 

\begin{figure}[t]
\centering
\includegraphics[height=14 cm]{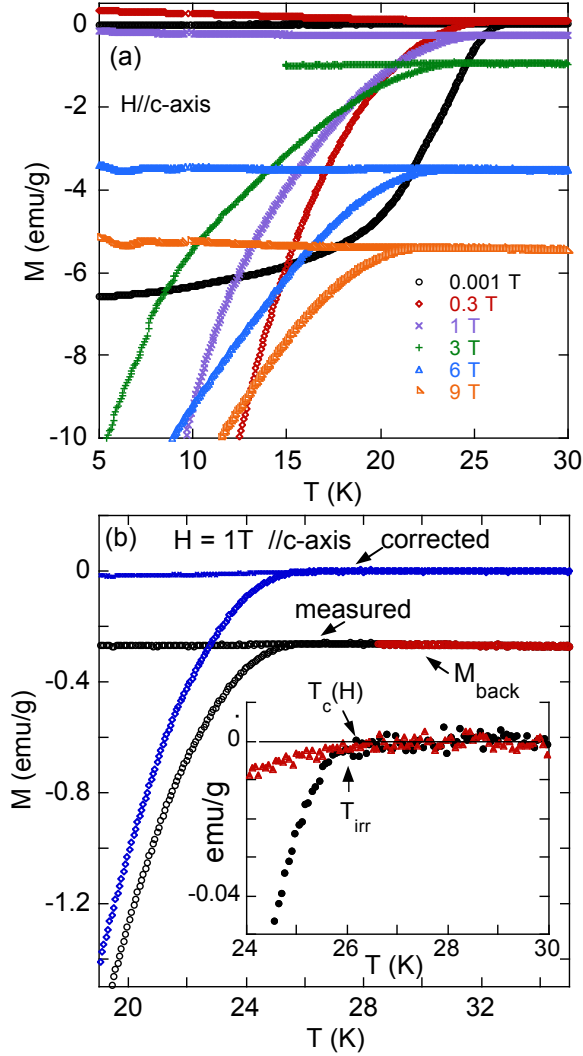}
\caption{(a) Selected isofield $M$($T$) curves as measured. (b) Isofield $M$($T$) curves with and without background subtraction for $H$ = 1 T. The inset shows an enlarged plot at temperatures near $T_c$(H).}
\label{fig1}
\end{figure}

\begin{figure}[t]
\centering
\includegraphics[height=14 cm]{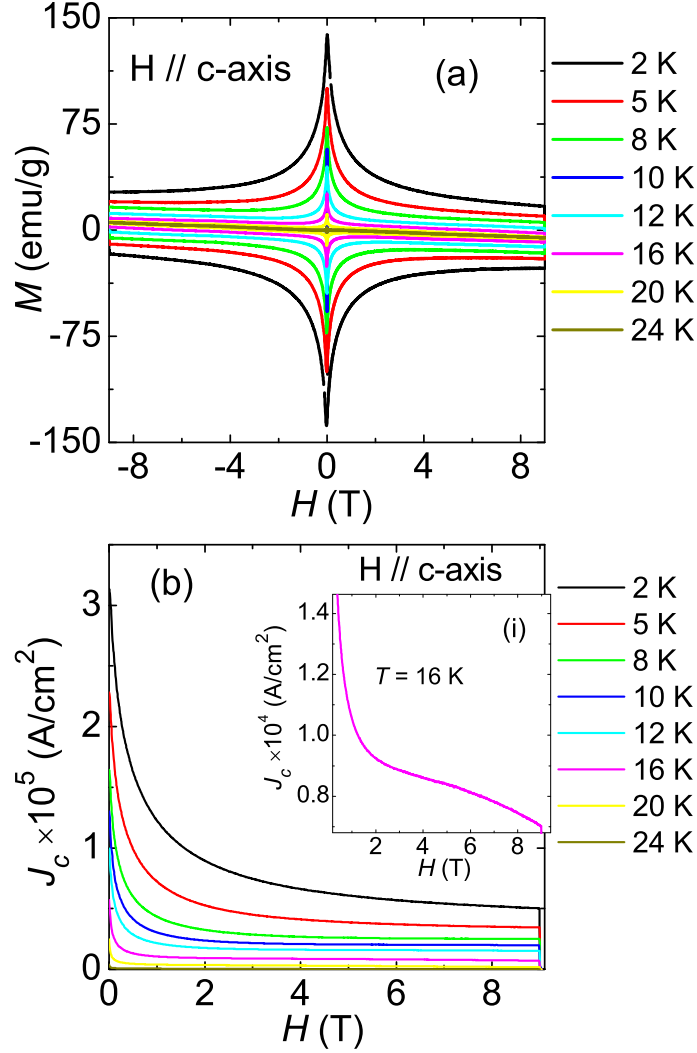}
\caption{(a) Isothermal $M$($H$) curves as measured. (b) Isothermal $J_c$($H$) curves as obtained from the Bean model. The inset shows detail of $J_c$($H$) curve at 16 K, showing a change in curvature as $J_c$($H$) decays.}
\label{fig2}
\end{figure}

Therefore, we decided to obtain the irreversible field, $H_{irr}$, from isothermal hysteresis  $M$($H$) curves as shown in Fig. 2a, which was limited to our maximum magnetic field of 9 T. It is possible to see from Fig. 2a that all $M$($H$) curves show a small asymmetry due to a rotation of the x-axis which is an effect due to the sample holder signal. This rotation can be fixed by calculating the equilibrium magnetization, $M_{eq}$ = ($M^+$+$M^-$)/2 where $M^+$ is the increasing field branch of a $M$($H$) and $M^-$ is the decreasing field branch. The new increasing and decreasing fields branches can then be obtained after subtracting the measured $M$($H$) from $M_{eq}$ which produces symmetric $M$($H$) curves with respect to the $x$-axis, evidencing that bulk pinning dominates in the sample. We define $H_{irr}$ as the magnetic field where the field increasing and field decreasing branches of the $M(H)$ curves merge together within the equipment resolution, $\sim$1x10$^{-5}$ emu. The critical current density associated to each $M$($H$) curve is obtained through the Bean's model \cite{bean} using expression \cite{ume87}, $J_c$ = 20$\Delta{M}$/$a$(1-$a$/3$b$) where $b$$>$$a$ (cm) are the single crystal dimensions defining the area perpendicular to the magnetic field, $\Delta{M}$(emu/cm$^3$) corresponds to the width of the hysteresis curves and the resulting $J_c$ is given in A/cm$^2$ units. Figure 2b shows the resulting $J_c$($H$) curves as obtained. The curves for  $T$$<$12 K show $J_c$(0) values, the critical current at zero field, above 10$^5$ A/cm$^2$ which is considered the required threshold value for applications \cite{pal15}. However, $J_c$ suppresses relatively fast as magnetic field increases and shows $J_c$ of the order of $\sim$10$^4$ A/cm$^2$ up to 9 T (max. field of measurements) for $T$ $<$ 12 K. We believe that the observed critical current density may be improved by introducing artificial defects as observed in other iron-pnictide superconductors \cite{tos16}. Among the different families of pnictide superconductors, the 122-class is the most relevant for high field application purposes \cite{pal15}. For comparison, in the case of K-doped (x = 0.3) 122-pnictide superconductor, the observed $J_c$ is over 10$^5$ A/cm$^2$ in a wide temperature and magnetic field range (up to 0.8$T_c$, 6 T) \cite{don16, shi17}. Whereas, in case of Co-doped (x = 0.057) and P-doped (x = 0.3) 122-pnictide superconductor, $J_c$ $>$ 10$^5$ A/cm$^2$ is observed for T = 0.5 $T_c$ and $H$ = 6 T \cite{shi17}. Similarly, near optimal doped Ni-122 pnictide superconductors shows $J_c$ $>$ 10$^5$ A/cm$^2$ in a wide temperature and magnetic field range \cite{ss19}. The effect of particle irradiation on the vortex dynamics and $J_c$ would be interesting to explore in the present sample. A detailed review of the critical current density and pinning in bulk, thin films, tapes and wires of IBS for technological importance is provided in Refs. \cite{pal15, hid18}. 

The inset of Fig. 2b shows a selected $J_c$($H$) curve at 16 K, evidencing a change in curvature as $J_c$($H$) approaches zero which was observed in all curves at higher temperatures. The same effect can be observed in the respective $M$($H$) curves. This change in curvature (downward to upward curvature as field increases) is usually a precursor of the peak effect occurring in $J_c$($H$) curves \cite{shyam}, which is absent in our sample. The interesting point of this effect is that $J_c$($H$) approaches zero with a downward curvature, while usually it approaches zero exponentially (upward curvature). 
 
\begin{figure}[t]
\centering
\includegraphics[height=15 cm]{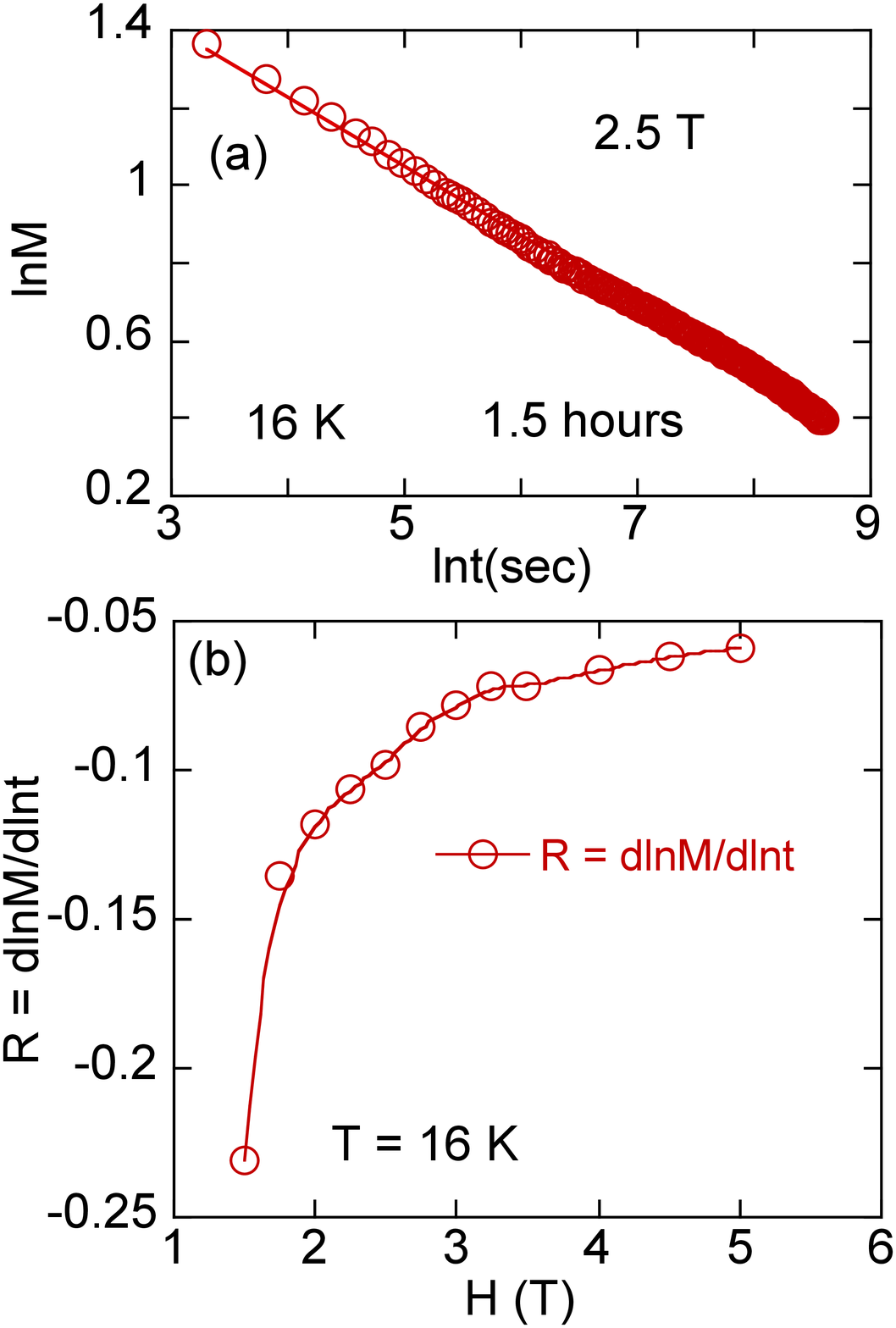}
\caption{(a) A typical magnetic relaxation curve measured during 1.5 hours, showing usual linear time dependence. (b) Relaxation rate, $R$  = $-$d[ln($M$)]/d[ln$t$], for $T$ = 16 K, plotted as a function of magnetic field.}
\label{fig3}
\end{figure}

In order to better understand this effect we have obtained magnetic relaxation measurements, $M$($t$), as a function of magnetic field and temperature. The resulting $M$($t$) curves plotted as ln($M$) vs. ln$t$ produced the usual linear behavior allowing us to obtain the relaxation rate $R$  = $-$d[ln($M$)]/d[ln$t$], as shown in Fig. 3a. Figure 3b shows a plot of $R$ vs $H$ at $T$ = 16 K, where $R$ decreases monotonically as field increases, evidencing the strength of the pinning as the field increases. 

\begin{figure}[t]
\centering
\includegraphics[height=15 cm]{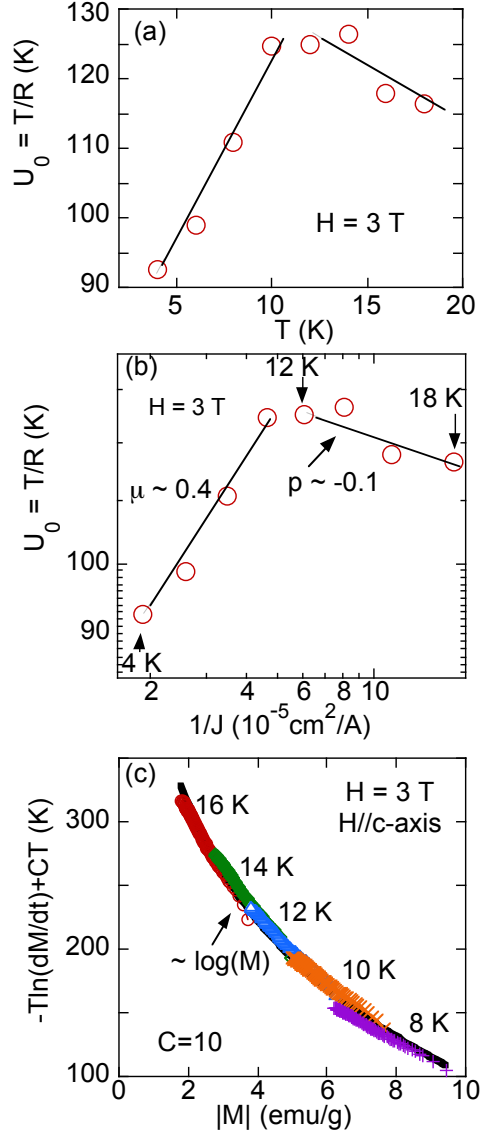}
\caption{(a) A plot of $U_0$ vs $T$ for $H$ = 3 T. (b) Apparent activation energy, $U_0$ obtained from magnetic relaxation for $H$ = 3 T, plotted against 1/$J_c$($T$). (c) Activation energy curve obtained from Maley's approach for $H$ = 3 T plotted against magnetization.}
\label{fig4}
\end{figure}

Figure 4a shows a plot of $U_{0}$ vs. $T$ for $H$ = 3 T. Two different slopes at intermediate temperature suggest a change in the pinning mechanism. Figure 4b shows a plot of the apparent activation pinning energy, $U_{0}$ = $R$/$T$, against 1/$J_c$, obtained for $H$ = 3 T, which despite the fact that the values of the exponents do not correspond to the expected ones \cite{wei} ($\mu$ $\sim$ 1 and $p$ $\sim$ -0.5), the overall behavior suggests the existence of a crossover in the pinning mechanism of the type elastic to plastic occurring at some temperature between 10 K and 14 K \cite{wei}. This crossover has been associated to the fish-tail effect found in $M$($H$) curves \cite{wei} which is absent in ours curves. Similarly, in K-doped BaFe$_2$As$_2$, a pinning crossover is suggested through $U_0$ vs. 1/$J$ plot, however, no fish-tail effect is observed for that temperature range \cite{shyam}. Therefore, it would be misleading to interpret the behaviour of $R$ vs. $T$ and $U_0$ vs. 1/$J$ plots as a pinning crossover, because, there is no effect of such crossover in the measured isothermal $M(H)$ curves. Figure 4c shows a plot of the activation energy $U$= -$T$ln(d$M$/d$t$)+$C$$T$ as obtained in Ref.\cite{maley}, where the smooth curve was obtained with a constant $C$=10. Parameter $C$ depends on the attempt frequency, hoping distance  and sample dimension. It must be mentioned that the smooth curve following a log($M$) behavior, as in Ref.\cite{maley} for YBaCuO, was obtained without the need of a temperature scaling. Also, the values of $U$ in Fig. 4c are of the same order of magnitude of the values found for YBaCuO in Ref.\cite{maley} at similar reduced temperatures. The scaling of $U$($M$) with magnetic field for data obtained at $T$ = 16 K, as performed in Refs.\cite{abulafia, FeCo, BaK}, would point to a possible pinning crossover, as suggested in Fig. 3b. Nevertheless, we could not find any reasonable scaling at this temperature. 

\begin{figure}[t]
\centering
\includegraphics[height=12 cm]{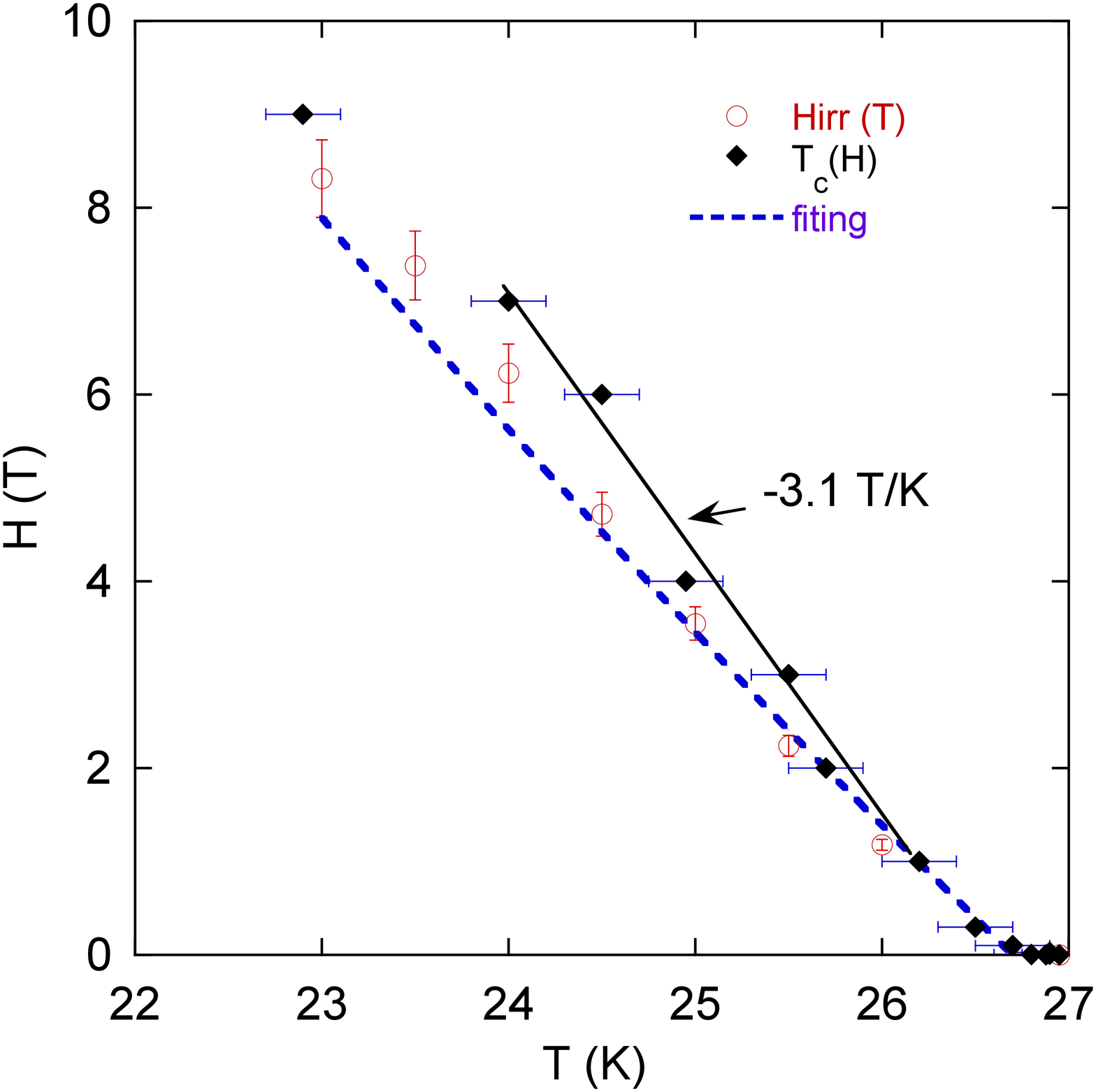}
\caption{Vortex phase-diagram of La$_{0.34}$Na$_{0.66}$Fe$_2$As$_2$. The dotted line is a fitting to the irreversibility line curve (see text for details).}
\label{fig5}
\end{figure}

Figure 5 shows the vortex phase-diagram where the values of $H_{irr}$ are very close to values of $T_c$($H$). Such a wide irreversible region, as shown in Fig. 2a, suggests that the studied system has potential for applications. In the phase-diagram (Fig. 5), at high temperatures, $T_c$($H$) increases linearly with d$H_{c2}$($T$)/d$T$ = -3.1 T/K, which from the Werthamer-Helfand-Hohenberg (WHH) formula \cite{whh}, renders a $H_{c2}$(0) = -0.695 $T_c$ d$H_{c2}$($T$)/d$T$ = 58 T. This value is higher than the Pauli paramagnetic limit for $H_{c2}$(0) given by $H_p$ = 1.84 $T_c$ =  50 T\cite{pauli}. The Maki parameter\cite{whh} defined as $\alpha$ = $H_{c2}$(0)/$\sqrt{2}$$H_p$ = 0.82 for the studied sample. Systems possessing $\alpha$$>$ 1 are candidates to exhibit the exotic Fulde-Ferrell-Larkin-Ovchinnikov (FFLO) phase\cite{fflo1,fflo2}. Since our system has an anisotropy $\gamma$= 1.9 \cite{1} the FFLO phase is a candidate to appear for $H$$\parallel$$ab$ planes. We observed that the irreversibility line does not follow a power law behavior with ($T_c$-$T$) as is commonly observed in many superconductors. In addition, the proximity of the irreversibility line to the $T_c$($H$) line resembles the behavior observed in low $T_c$ systems, as for instance in NbSe$_2$ \cite{nbse}. $H_{irr}(T)$ has been described in Refs. \cite{baruch0, baruch1} by an expression which takes in to account the disorder in the system. Even for NbSe$_2$ the $H_{irr}(T)$ line is visibly further apart from the $H_{c2}(T)$ line than what is observed in our sample. The dotted line in Fig. 5, represents the best fit of the $H_{irr}(T)$ data to the expression provided in Refs. \cite{baruch0, baruch1}, which is mentioned below.

1-$t$-$b$+3$n_p$(1-$t$)$^2$ 4$\pi$-2 $\sqrt{2G_i}$ $t$$b$=0

where $t$ = $T$/$T_c$ is the reduced temperature, $b$ = $H$/$H_{c2}$(0), $n_p$ measures the disorder in the system and $G_i$ is a different definition of the Ginzburg number which measures the importance of thermal fluctuations. In that expression, $b$, $n_p$ and $G_i$ are fitting parameters, and the dotted line shown in Fig.5 was obtained for a slightly higher $T_c$ = 28 K, $n_p$ = 0.002, $G_i$$\sim$10$^{-8}$ and $H_{c2}$(0) = 60 T. It is observed that both $G_i$ and $n_p$ show a good fit within 10 \% of the fitted values, whereas, $H_{c2}$(0) may vary within 4\% of the value obtained from the fitting. The value of $n_p$ is similar to that obtained in Ref. \cite{baruch0} for NbSe$_2$, $H_{c2}$(0) is similar to the value obtained from the WHH expression, but $G_i$ is three orders of magnitude lower than the value obtained for NbSe$_2$. We do not have explanation for such a low value of $G_i$ in the present case. The importance of thermal fluctuations is sometimes associated to the extension of the reversible region, which in our case is virtually absent from the $M(T)$ curves. 

\begin{figure}[t]
\centering
\includegraphics[height=10 cm]{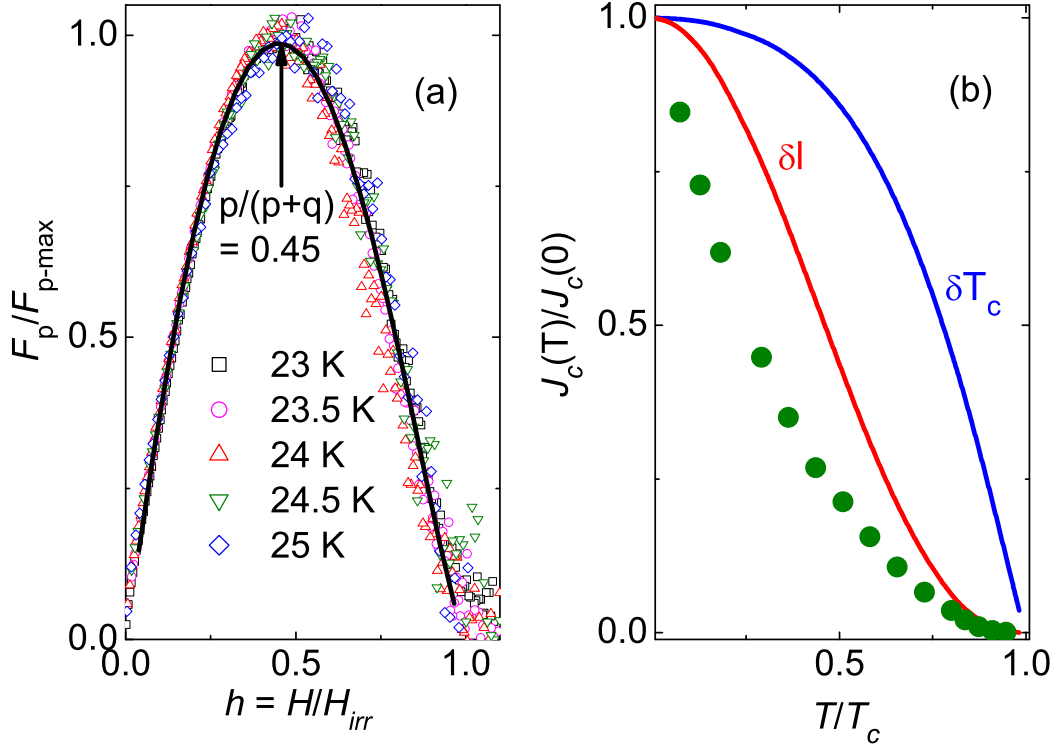}
\caption{(a) Normalized pinning force density plotted against the reduced magnetic field. The experimental data taken at different temperatures collapsed into a single scaled curve, which is fitted using the expression, $F_p$/$F_p$$_{max}$ = $A$$h^p$(1-$h$)$^q$, where, $p$ and $q$ are fitting parameters, which define the pinning characteristics. (b) Normalized critical current density, $J_c$($T$)/$J_c$(0) as a function of reduced temperature, $T$/$T_c$. The solid lines represents the $\delta l$ and $\delta T_c$ pinning behavior.}
\label{fig6}
\end{figure} 

We present in Fig. 6a, the plot of the normalized pinning force density, $F_p$/$F_p$$_{max}$ as a function of the reduced magnetic field $h$=$H$/$H_{irr}$, where, $F_p$=$J_c$$\times$$H$. The collapse of the many different isothermal pinning force curves is evident on the graph. This plot allows us to obtain the reduced magnetic field ($h$) for which the pinning force reaches its maximum, $h_{max}$ = 0.45, and a fitting of the resulting curve to the Dew-Hughes formula \cite{dew} $F_p$/$F_p$$_{max}$ = $A$$h^p$(1-$h$)$^q$ produced $A$ = 5.14, $p$ = 1.08 and $q$ = 1.32, where $p$/($p$+$q$) = $h_{max}$ = 0.45 as obtained from the experimental data. This fitting has been applied to examine the pinning force on many systems over the years\cite{kobli} and the different values obtained for the parameters $h_{max}$, $p$ and $q$, are used to determine the type of the dominant pinning\cite{dew, kobli, mat13, sha13}. In a classic paper, Dew-Hughes discussed the various scenarios of pinning centers involved in different pinning mechanisms \cite{dew}. For a system having pinning due to the variation in charge carrier mean free path ($\delta$$l$-pinning) and the pinning centers are point-like, the maximum in the normalized  pinning force density occurs at $h$ = 0.33, with $p$ = 1 and $q$ = 2. In the case of the pinning due to the variation in the superconducting transition temperature ($\delta$$T_c$-pinning), the maximum in the normalized pinning force density is found to be at much higher $h$ values, in short, for point pinning centers, $h$ = 0.67, with $p$ = 1, $q$ = 2; for surface pinning centers, $h$ = 0.6, with $p$ = 1.5 and $q$ = 1 and for volume pinning centers, $h$ = 0.5, with $p$ = $q$ =1. Therefore, in the present study, $h$ = 0.45 with $p$ = 1.08 and $q$ = 1.32 shows that a single type of pinning is not sufficient to explain the results. However, similar values of $h$ has also been observed in many other studies of iron-pnictide superconductors, such as, for BaFe$_{1.9}$Ni$_{0.1}$As$_2$, $h$ = 0.4 \cite{sha13}, for Ca$_{0.8}$La$_{0.2}$Fe$_{1-x}$Co$_x$As$_2$, $h$ = 0.44 \cite{wei}, for Ba$_{0.68}$K$_{0.32}$Fe$_2$As$_2$, h = 0.43 \cite{sun09}, for Ba(Fe$_{1-x}$Co$_x$)$_2$As$_2$, $h$ = 0.45 \cite{yam09}. In such studies, the value of $h$ $\approx$ 0.45, is attributed to an inhomogeneous distribution of dopants or Arsenic deficiency \cite{sun09}. Shahbazi et al \cite{sha13} argued that $h$ = 0.4, in case of BaFe$_{1.9}$Ni$_{0.1}$As$_2$ is due to the $\delta$$l$-type pinning. Zhou et al \cite{wei} related $h$ = 0.44 with the randomly distributed nano-scale point-like defects, which is common in the case of iron pnictide superconductors \cite{hua08, sun09, wei}. Therefore, $h$ = 0.45 in our case is attributed to the $\delta$$l$-type pinning due to point pinning centers. Whereas, for $\delta$$T_c$ pinning, the maximum in normalized pinning force density would occur at $h$ higher than 0.5. The type of pinning can be further examined following an approach developed in Ref.\cite{wen} where the temperature dependent critical current at zero field $J_c$($T$), normalized by the critical current at zero field at $T$ = 0 is plotted against $t$ = $T$/$T_c$. The equivalent plot for our data is shown in Fig. 6b, which is compared to the theoretical expression for $\delta$$l$-type of pinning, where $J_c$($T$)/$J_c$(0) = (1 + $t^2$)$^{-1/2}$(1- $t^2$)$^{5/2}$ and for $\delta$$T_c$-type pinning, $J_c$($T$)/$J_c$(0) = (1-$t^2$)$^{7/6}$(1 + $t^2$)$^{5/6}$ \cite{wen}. It is evident from Fig. 6b that the $\delta$$l$-type pinning alone can not explain the experimental data adequately. 

\section{Conclusions}

In conclusion, the vortex phase-diagram of the newly synthesized iron-based superconductor La$_{0.34}$Na$_{0.66}$Fe$_2$As$_2$ shows an irreversibility line very close to the mean field transition temperature $T_c$($H$) evidencing a strong pinning. The irreversibility line does not follow the usual power law with ($T_c$-$T$) but it was successfully fitted by an expression developed in Refs. \cite{baruch0, baruch1} considering the effect of disorder, where a considerable low disorder similar to that observed for NbSe$_2$ \cite{baruch0, baruch1} was found for our system. We observed that the upper critical field at zero temperature exceeds the prediction of the Pauli paramagnetic limit, suggesting that the system is a candidate to show the FFLO phase for $H$$\|$$ab$-planes. The critical current density at zero magnetic field reaches the threshold value $J_c$$>$10$^5$ A/cm$^2$ for temperatures below 12 K, which along with the fact that the irreversibility line is very close to $T_c$($H$) makes the system technologically relevant. Magnetic relaxation obtained as a function of field for a fixed temperature shows that the relaxation rate monotonically decreases as the field increases, while magnetic relaxation obtained for a fixed field as a function of temperature suggests a crossover in the pinning mechanism. The latter results allowed us to obtain a smooth curve of the isofield activation energy with magnetization, as first done by Maley\cite{maley}, where the observed log($M$) behavior and values are similar to the ones obtained in Ref. \cite{maley} for YBaCuO. The pinning analysis using the Dew-Hughes model suggests a $\delta$$l$-type pinning due to the point pinning centers. However, the temperature dependence of the critical current density indicates that $\delta$$l$ pinning alone can not explain the data adequately. To explore this compound for technological use, the effect of grain boundaries on the critical current density and vortex dynamics in polycrystalline and thin film samples are yet to be investigated.

\section*{Acknowledgments}

SS acknowledges financial support from a post doctoral fellowship by Fundação Carlos Chagas Filho de Amparo à Pesquisa do Estado do Rio de Janeiro - FAPERJ (Project E-26/202.848/2016). LG was also supported by FAPERJ (Projects E-26/202.820/2018 and E-26/010.003026/2014). SSS and ADA were partially supported by the Brazilian federal agency Conselho Nacional de Desenvolvimento Científico e Tecnológico - CNPq. The work at IOP, CAS is supported by the National Key R$\&$D Program of China (Nos. 2017YFA0302900, 2016YFA0300502, 2017YFA0303103), the National Natural Science Foundation of China (Nos. 11674406, 11674372, 11774399), the Strategic Priority Research Program (B) of the Chinese Academy of Sciences (XDB25000000, XDB07020000). HL is supported by the Youth Innovation Promotion Association of CAS (No. 2016004). 

\section*{References}

\end{document}